\documentclass[useAMS,usenatbib]{mn2e}
\usepackage{hyperref}
\usepackage{amssymb,esint}
\usepackage{graphicx}
\usepackage{epstopdf}
\usepackage{mathptmx}
\newcommand{\bea}{\begin{eqnarray}}
\newcommand{\eea}{\end{eqnarray}}
\newcommand{\beq}{\begin{equation}}
\newcommand{\eeq}{\end{equation}}

\title[Comments on arXiv:1310.5067]{Comments on arXiv:1310.5067:  ``A response to `A self-consistent public catalogue of voids and superclusters in the SDSS Data Release 7 galaxy surveys' " }
\author[S. Nadathur \& S. Hotchkiss]{Seshadri Nadathur$^{1,3}$ \& Shaun Hotchkiss$^{2,1}$\\
$^1$Department of Physics, University of Helsinki and Helsinki Institute of Physics, P.O. Box 64, FIN-00014, University of Helsinki, Finland\\
$^2$Department of Physics and Astronomy, University of Sussex, Brighton, BN1 9QH, UK\\
$^3$Fakult\"at f\"ur Physik, Universit\"at Bielefeld, Postfach 100131, D-33501 Bielefeld, Germany}

\begin{document}

\date{\today}

\pagerange{\pageref{firstpage}--\pageref{lastpage}}

\label{firstpage}

\maketitle

\begin{abstract}
\citeauthor{Sutter:2013resp} have responded to the criticisms we made of their cosmic void catalogue in our recent paper presenting an alternative catalogue \citep{Nadathur:2013bba}. Unfortunately, their response contains several statements which are incorrect, as we point out in this note.
\end{abstract}

\label{firstpage}

\maketitle

In \citet{Nadathur:2013bba} (hereafter NH13) we recently discussed the creation of a self-consistent public catalogue of cosmic voids and superclusters identified in the Sloan Digital Sky Survey (SDSS) Data Release 7 galaxy redshift surveys through the use of a watershed-transform structure-finding algorithm based on {\small ZOBOV} \citep{Neyrinck:2007gy}. We embarked on the creation of such a catalogue because of the discovery of serious errors and inconsistencies in an existing catalogue of voids  (\citealt{Sutter:2012wh}; hereafter S12), and our paper contains a critical discussion of this catalogue as well as our own new results. 

Sutter et al. have responded to our criticisms in their own note (\citealt{Sutter:2013resp}; hereafter S13). Unfortunately however, this response itself contains a number of incorrect or misleading statements.  In this short note we address some of these.

The first and most important point to clarify is the nature of our criticism of the S12 catalogue. S13 state that our principal criticism \emph{``is that the regions identified as voids are not, on the whole, underdense"}. This is not correct. While we do find that several of the `voids' in the S12 catalogue have average densities $\rho_\rmn{void}$ up to 10 times larger then the mean density in the galaxy sample---which, incidentally, is \emph{not} the case for our voids---this is actually a symptom of other failings in the S12 methodology, rather than the primary flaw with the catalogue. 

In fact the more serious problem with the S12 catalogue lies in the method used to reconstruct the density field using  the Voronoi tessellation estimator, in particular the use of boundary mock particles intended to handle survey edge effects. As we have explained in NH13, for three of the galaxy samples studied, the density of boundary mocks used by S12 is \emph{ lower} than that of real galaxies, and thus completely insufficient to prevent galaxy Voronoi cells from leaking out of the surveyed volume, resulting in them being assigned artificially low densities. Also of concern is the fact that no mocks are placed above the maximum redshift extent of the survey, which means that Voronoi cells of galaxies at high redshifts can and do leak out of the survey volume. This results in incorrect assignment of density minima. These problems have  been explained in detail in NH13, where we have also discussed other issues, such as the absence of correction for a varying selection function and the use of an incorrect survey mask, which also affect the reconstruction of the density field.

Since the correct reconstruction of the density field is the necessary starting point for the rest of the void-finding algorithm, these problems with the S12 methodology are the primary concern. There are also several secondary issues regarding inconsistent and inappropriate use of void selection criteria, some of which we address again below. However, we wish to emphasise that these are not the reason our catalogue in NH13 has a completely different character to that in S12. In particular,  contrary to the implicit impression conveyed by S13, applying a stricter cut on minimum density or minimum radius (or any other variable) on the `voids' in the S12 catalogue \emph{does not} recover our catalogue as a special case. 

Having thus clarified the overall nature of our criticism of the S12 catalogue and approach, we turn to some of the details of the S13 response.

\section{`Void' densities}

\subsection{Mean densities}
In stating what they believed to be our principal criticism of their paper, S13 have in fact misquoted our paper. The correct statement is actually that \emph{``the identified void locations do not, on average, correspond to underdense regions in the galaxy density field"} (p.2 of NH13). This is a reference to the fact that for most samples the average radial profile of all `voids', stacked around their barycentres, does not show significant underdensities at the location of the  barycentre. This was clearly shown in Figure 10 of NH13, and can be contrasted with the equivalent figure for our voids (Figure 9).

It is also misleading to claim, as S13 do, that the inclusion of `voids' which are highly overdense was ``by design". It is true that due to the inclusion of high-density galaxies at void edges, {\small ZOBOV} is known to report voids with average densities biased high, and our catalogue shows this behaviour too (though with nowhere near the severity of the S12 catalogue). This is why we argue that the average density is not an appropriate criterion for void selection. However, the published version of the S12 paper states clearly that the authors had modified the {\small ZOBOV} algorithm to report only those `voids' with mean density contrasts of $\delta\leq-0.8$, quite the opposite of the implication of the ``by design" statement.

In fact, it has subsequently emerged\footnote{see the updated statement at  \url{http://www.cosmicvoids.net/updates/anoteaboutdensitythresholds}} that the authors, while intending to apply a cut on the overall density, were in fact implementing a completely different and unintended criterion, which restricted the merging of density minima zones to form voids according to the watershed algorithm. According to the statement on their website, this criterion---if correctly applied---should in fact have been similar to but substantially more lenient than the merging criterion that we apply for the creation of Type2 voids. This means that merging of zones to form voids should occur \emph{more} frequently in the S12 case than for our Type2 voids. 

However, whereas our Type2 voids often contain multiple merged zones, \emph{all} `voids' in the S12 catalogue, without exception, consist of only a single zone each. It appears that Sutter et al.'s latest explanation of how their algorithm works is still inconsistent with their catalogue. Preventing \emph{any} merging of zones to form voids seems to defeat the entire point of using the watershed transform algorithm. It also represents maximal interference in the void-finding operation, exactly the opposite of the philosophy now claimed in S13.

\begin{figure}
\includegraphics[width=83mm]{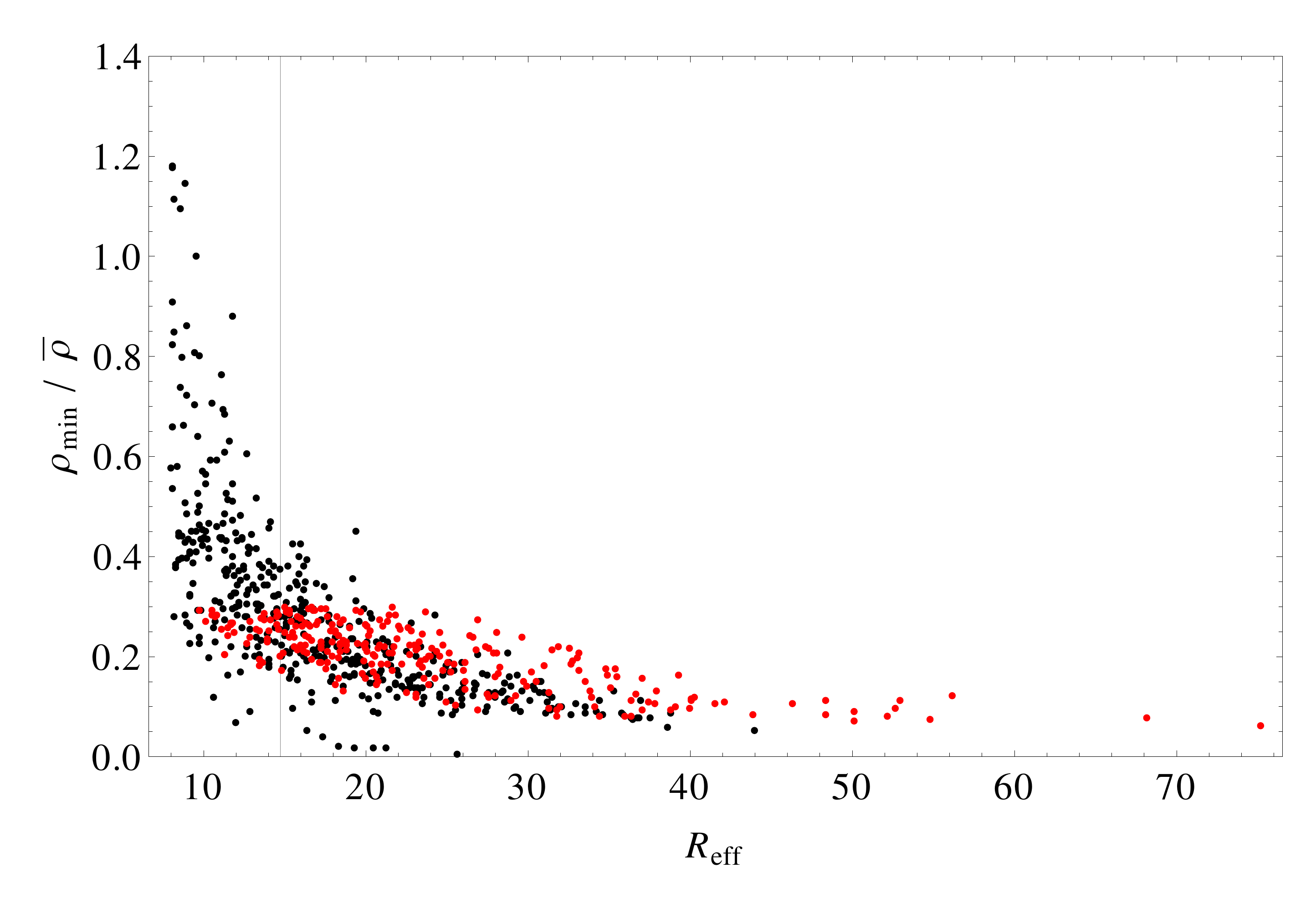}
\caption{Minimum void density versus effective radius for voids in the \emph{bright1} sample. Black dots are for `voids' in the S12 catalogue, red points are for Type1 voids in NH13. Note that Type1 voids are selected to have $\rho_\rmn{min}\leq0.3\overline{\rho}$. The vertical line is the minimum radius cut suggested by S13.} 
\label{figure:F1}
\end{figure}

\subsection{Central densities}

S12 claim to apply a central density cut to reject from their catalogue any void which contains a region of density contrast $\delta>-0.8$ within a defined central region, yet even according to their own Voronoi-reconstructed density field the majority of their `voids' have \emph{minimum} $\delta_\rmn{min}>-0.8$. The explanation for this lies in the design of the central density cut they apply: this is based not on the reconstructed density field, but on taking a sphere of radius one-quarter the `void' effective radius $R_\rmn{eff}$, and simply counting the number of galaxies within it.

To better understand the problem with this method, let us take a concrete example. Consider a void in the \emph{dim1} sample, with an effective radius $R_\rmn{eff}=7\;h^{-1}$Mpc, which is larger than both the median reported value for the S12 catalogue, and the new minimum effective radius cut that S13 now recommend. The overall mean number density of galaxies in the \emph{dim1} sample is only $2.4\times10^{-2}\;(h^{-1}\rmn{Mpc})^{-3}$, corresponding to on average roughly one galaxy in every $42\;h^{-3}\rmn{Mpc}^3$. The radius of the central sphere for our example void is $0.25\times7=1.75\;h^{-1}$Mpc, and its volume is roughly $22\;h^{-3}\rmn{Mpc}^3$. A sphere of this size, if placed at completely random locations in the galaxy distribution, will contain no galaxies at all a very significant fraction of the time. It would then pass the S12 central density cut, even though the Voronoi-reconstructed density field at that point may be far from zero. Even worse, the sphere will quite often by chance contain a single galaxy; since this is almost twice the `expected' number, it will then fail the S12 cut, even if the region actually happens to be genuinely underdense.

In effect then, the central density selection cut applied by S12  is so badly affected by Poisson noise as to be essentially indistinguishable from it. It is far preferable to use the less noisy Voronoi tessellation information, which is after all available and already used by {\small ZOBOV}. 

\subsection{Density profiles}

Sutter et al. claim that the stacked profiles in Figure 7 of S12 show that their `voids' are underdense near the centre and do show a qualitatively universal behaviour. However, closer inspection of this figure reveals that each stack contains only a small specially chosen subset of `voids' in each sample. Clearly the behaviour seen in these particular subsets cannot be universal: if it were then the behaviour would be reproduced when the stacks are extended to include all `voids', and Figure 10 of NH13 shows that this is not the case.

We also note in this context the claim in S13 that the universal behaviour may not be reproduced because some of their `voids' have higher compensation regions at the walls and \emph{also} have lower density contrasts (this is the quantity we refer to as the \emph{density ratio}, $r$). The obvious corollary of these two statements and the definition of the density ratio is that these `voids' do not have underdense centres, i.e. they are not, in fact, voids.

\section{Minimum void radius}

S13 state that although in their catalogue they keep all `voids' with $R_\rmn{eff}\geq R_\rmn{eff,min}=n_g^{-1/3}$, they have always cautioned users to not trust the smallest voids, and they now recommend a cut of $R_\rmn{eff}\geq 2R_\rmn{eff,min}$ instead. We find this statement rather disingenous. While the authors' website does contain a warning about small voids, they do not appear to have heeded their own advice: several recent papers to which some or all of the original authors have contributed \citep[e.g.][]{Sutter:2012tf,Hamaus:2013,Melchior:2013,Sutter:2013ssy} have made use of the S12 catalogue, and not one of them has imposed such a minimum radius cutoff.

In any case, not only is the new choice of a factor of 2 in the recommended radius cutoff completely arbitrary, it actually does not even solve the problem. Not only do many `voids' with $R_\rmn{eff}\geq 2R_\rmn{eff,min}$ still have $\rho_\rmn{min}>0.2\overline\rho$, in contradiction with the intention of the S12 central density cut, but such a minimum radius cut would also unnecessarily exclude any genuine voids that might happen to have smaller $R_\rmn{eff}$. To illustrate this point, in Figure~\ref{figure:F1} we show the distribution of void minimum densities and radii as in Figure 1 of S13, with our own voids from the same sample overlaid on top. This figure also clearly shows that our catalogue is not a special case of the S12 one.

\section{Other items}

\emph{Statistical significance of voids}: S13 imply that the voids in our catalogue fail the test of significant difference from those found due to Poisson noise in the same way that theirs do. This is incorrect. All our voids are constructed to be statistically significant, as explained in NH13. \citet{Neyrinck:2007gy} discusses one method of judging this significance, based on the void density ratio alone. We showed that voids can also be judged to be distinct from noise on the basis of their $\rho_\rmn{min}$ values. Our criticism of the S12 catalogue is that $40\%$ of their `voids' pass \emph{neither} test.

\emph{Coordinate system}: S13 also claim that studies such as those of \citet{Planck:ISW} and \citet{Melchior:2013} confirm via measurement of gravitational effects that the voids in their redshift-coordinate catalogue capture physical underdensities. However, the actual result of the \emph{Planck} team's search for the ISW signal of voids was that after stacking a variable number of voids that was allowed to range all the way from 1 to $1985$, the detection of the signal ranged \emph{``from negligible to $2.5\sigma$ at best"} \citep{Planck:ISW}, \emph{without} accounting for any `look-elsewhere' effect. This is hardly the strong corroboration claimed by S13.

In our opinion, the results of \citet{Melchior:2013} suffer from a similar statistical ambivalence. In any case, both these analyses rest crucially on the assumption that the `voids' in the S12 catalogue can be at least approximately modelled as spheres of their effective radii. That this is not true is amply demonstrated by the stacked radial profiles in Figure 10 of NH13. 

\emph{Survey boundary contamination}: In an earlier verion of NH13, and of this note, we stated that a large number of the core or minimum-density galaxies of the S12 reported voids were adjacent to the boundary mocks used by S12. This statement was incorrect, since we were inadvertently using additional mocks as well. We would therefore like to amend our statement. The correct statement is that for $\sim35\%$ of all the structures listed in S12, the minimum-density or core galaxy about which the structure is based is adjacent to a mock particle when mock particles are placed above the maximum redshift extent (to prevent leakage in the radial direction) and have a number density sufficient to prevent leakage in other directions. 

This means that these structures were based around galaxies whose Voronoi cells---with the S12 placement of mocks---were leaking out of the surveyed volume and were therefore assigned artificially low densities, but were not being correctly identified as edge-contaminated. If they had been so identified, they would not have constituted density minima, and therefore would not have formed voids. For the lower-redshift main galaxy samples, the insufficient density of S12 mocks is the more important factor in this edge contamination, while for the LRG samples it is the absence of high-redshift mocks. The discussion of this point has been appropriately modified in Section 4.2 of NH13, including the addition of an extra figure (Figure 2).

\section*{Acknowledgements}
SN is supported by Academy of Finland grant 1263714. SH is supported by the Science and Technology Facilities Council (grant number ST/I000976/1). 

\bibliography{../../refs.bib}
\bibliographystyle{mn2e}

\label{lastpage}
\end{document}